\documentclass[conference]{IEEEtran}
\usepackage[margin=1in]{geometry}
\usepackage{comment}
\usepackage{graphicx}
\usepackage{url}
\usepackage{todonotes}
\usepackage{multirow}
\usepackage{footnote}
\usepackage{matlab-prettifier}
\usepackage{listings}
\usepackage{flushend}

\makesavenoteenv{tabular}
\makesavenoteenv{table*}

\title{High-throughput Ingest of Data Provenance Records into Accumulo}
\author{\IEEEauthorblockN{Thomas Moyer}
\IEEEauthorblockA{MIT Lincoln Laboratory\\
Lexington, MA 02420\\
Email: tmoyer@ll.mit.edu}
\and
\IEEEauthorblockN{Vijay Gadepally}
\IEEEauthorblockA{MIT Lincoln Laboratory\\
Lexington, MA 02420\\
Email: vijayg@ll.mit.edu}
\thanks{DISTRIBUTION STATEMENT A. Approved for public release: distribution unlimited.}

\thanks{This material is based upon work supported under Air Force Contract No. FA8721-05-C-0002 and/or FA8702-15-D-0001. Any opinions, findings, conclusions or recommendations expressed in this material are those of the author(s) and do not necessarily reflect the views of the U.S. Air Force.}

\thanks{Delivered to the U.S. Government with Unlimited Rights, as defined in DFARS Part 252.227-7013 or 7014 (Feb 2014). Notwithstanding any copyright notice, U.S. Government rights in this work are defined by DFARS 252.227-7013 or DFARS 252.227-7014 as detailed above. Use of this work other than as specifically authorized by the U.S. Government may violate any copyrights that exist in this work.}}

\makeatletter
\def\lst@makecaption{%
  \def\@captype{table}%
  \@makecaption
}
\makeatother

\IEEEoverridecommandlockouts

\begin{document}

\maketitle

\begin{abstract}
  Whole-system data provenance provides deep insight into the processing of data on a system, including detecting data integrity attacks. The downside to systems that collect whole-system data provenance is the sheer volume of data that is generated under many heavy workloads. In order to make provenance metadata useful, it must be stored somewhere where it can be queried. This problem becomes even more challenging when considering a network of provenance-aware machines all collecting this metadata. In this paper, we investigate the use of D4M and Accumulo to support high-throughput data ingest of whole-system provenance data. We find that we are able to ingest 3,970 graph components per second. Centrally storing the provenance metadata allows us to build systems that can detect and respond to data integrity attacks that are captured by the provenance system.
\end{abstract}

\section{Introduction} \label{sec:intro}

Data provenance provides a history of the data as it is processed. This history has a variety of uses, including protecting against malicious changes\cite{db-rollback}, and detecting attacks that occur on the system\cite{prov-ids}. Many of the use cases for provenance require collection of large volumes of information, such as from whole-system provenance collectors, e.g. the Linux Provenance Modules, LPM~\cite{lpm}, or Hifi~\cite{hifi}.

While these whole-provenance systems ensure that a complete provenance record is collected, they introduce challenges for storage and analysis of the data, namely the volume of data generated. For storage, LPM and Hifi store data in flat files on disk, and the Provenance-aware Storage System, PASS~\cite{pass}, stores data in a BerkleyDB, also local to the system that is collecting the provenance. To give a sense of the volume of data, a system using LPM to collect provenance generates 2.5GB of provenance data during kernel compilation, an I/O-intensive workload. In order to do any processing, especially for distributed systems that are all collecting provenance, the data must first be moved to a central location. This often limits provenance to offline analysis, due to the sheer size of the data. What is needed is a system that can ingest provenance in a centralized location and support queries on that data.

In this paper, we leverage existing work on high-speed ingest for the Accumulo database~\cite{accumulo,d4m-ingest-paper}. We also leverage the D4M schema to support parallel processing and analytics~\cite{d4m,gadepally2015d4m}. Finally, we describe an analytic using these tools to identify the input files that the system uses to generate a given output file. Graph analytics are useful in analyzing provenance data, and recent tools have shown that server-side analytics are possible. For analyzing large volumes of provenance data stored in Accumulo, we can leverage existing tools, including Graphulo~\cite{graphulo,gadepally2015graphulo}. The first step in analyzing this data is to store it somewhere amenable to analysis.

While many existing solutions for whole-system data provenance have focused on the problem of collection, they often fail to address storage issues, often relying on simple flat-files to store the collected data. The proposed solutions either focus on a single system, and assume a model where the provenance data will be post-processed on another system at a later time~\cite{pass,hifi}, or rely on in-memory solutions to store the data locally. Our system provides an online capability that supports analysis of provenance data from multiple machines in a centralized manner.

We build the ingest system using Accumulo and D4M, and characterize its performance under a variety of scenarios. We find that with a single ingest process running, our system can process 3,970 graph components (nodes or edges) per second. We then characterize the performance of the analytic process. The analysis identifies the set of inputs used to generate a specific output, e.g. the set of source files used to create a specific binary.

One limitation of whole-system provenance is the storage overhead and the need to ingest large volumes of provenance metadata quickly, in order to support online analytics. Our system provides an example of how to manage the tradeoff between performance and storage for whole-system provenance. The rest of this paper is organized as follows. Section~\ref{sec:background} provides more information on data provenance. Section~\ref{sec:design} details the design of our ingest pipeline. Section~\ref{sec:eval} presents a performance evaluation of our system. Finally, Section~\ref{sec:discussion} and~\ref{sec:conclusion} is a discussion of future work and conclusions, respectively.


\section{Background} \label{sec:background}

Data provenance is the history of data as it is processed. Provenance tracks metadata about the inputs and outputs to a process, the specific process, and the controlling agents of a process. The data can be collected at different levels of granularity within a system, including within applications, where developers have the most context or within the operating system, which provides complete coverage at the expense of relevant context. However, the limitation of application provenance is the required effort by the developers to instrument the application. An alternative is to run the application on an operating system that is provenance-aware, such as the Hifi~\cite{hifi}, or Linux Provenance Modules~\cite{lpm}. Such systems have been shown to provide complete provenance records, but suffer from high storage overheads.

From the provenance data, it is possible to reconstruct the history of a process, in order to determine the flow of data through a system. This history provides the ability to judge the authenticity of data. In addition, provenance has been demonstrated as a tool for attack detection~\cite{prov-attack-detect}, data loss prevention, and forensic analysis~\cite{prov-forensic-analysis}. These applications of provenance require an entire pipeline in order to leverage data provenance. The provenance pipeline can be broadly broken into several stages, described below. There are choices at each stage, and many of the choices directly impact the performance and utility of the collected provenance data.

\paragraph{Granularity} There is a trade-off between collecting everything at a low-level, e.g. whole-system provenance~\cite{lpm,hifi,pass} or collecting at a high-level, e.g. application provenance~\cite{prov-lib1,prov-lib2,prov-lib3,provtoolbox}. With low-level provenance, the collection is guaranteed to be complete, but lacking in context (i.e. a semantic gap), while high-level collection will be context-rich but is not guaranteed to be complete.

\paragraph{Collection} With the chosen granularity, a collection mechanism is chosen. For example, fine-grained collection (i.e. low-level) is achieved with any of the whole-system provenance systems such as LPM or Hifi. Context-rich high-level collection is achieved via any of the readily available provenance libraries~\cite{prov-lib1,prov-lib2,prov-lib3,provtoolbox}.

\paragraph{Encoding} Collected data must be encoded in a way that analytic tools can process the data. For interoperability, it is best to rely on the standard encoding, such as those defined by the W3C in the PROV specification~\cite{w3c-prov}. This encoding must also accurately reflect the processing done on the data, and this modeling of data processing can be challenging.

\paragraph{Storage} There are many choices for storing provenance data, including graph databases, like Neo4J~\cite{neo4j} and Titan~\cite{titan}, traditional SQL databases, or key-value stores, such as Cassandra~\cite{cassandra} and Accumulo. This paper focuses on the Accumulo database, due to the high throughput ingest~\cite{d4m-ingest-paper}, as well as recent efforts to provide increased security for the data stored in these databases~\cite{pace-summit,cmd}.

\paragraph{Analytics} Once provenance data is stored, analytics run to make use of the provenance data. For example, the provenance data can be analyzed to determine if a particular output is derived from a specific input that may be known to be malicious. The result of the analytics determine what response actions the provenance system takes.

\paragraph{Response} Responses from the provenance system can be passive, such as displays that provide information about the data, or active, by triggering corrective actions within the system. One example of a corrective action would be to quarantine any output that is derived from a malicious input.

\paragraph{Security} One over-arching consideration for the provenance data is security. The choices for security are determined by the choices made for each stage of the pipeline, and are too varied to include here. For whole-system provenance security, the reader is referred to~\cite{lpm} which details the security of the LPM system. Each stage above has security considerations that must be carefully considered, a problem that is outside the scope of this paper.

\begin{table}[t]
  \centering
  \begin{tabular}{|l|r|}
    \hline
    Event & Occurrences \\
    \hline
    \hline
    boot & 1 \\
    credfork & 336,505 \\
    exec & 47,475 \\
    fperm & 3,851,401 \\
    setid & 47,691 \\
    \hline
    \hline
    Total & 4,283,073
  \end{tabular}
  \caption{Each relevant LPM event is translated into at most two graph components, namely a node, and an edge. For a full explanation of the events that LPM captures, we refer the readers to~\cite{lpm}. The number of unique events that must be processed to build the full provenance graph exceeds 4,000,000 during approximately 38 minutes.}
  \label{tab:lpm-events}
\end{table}

For many provenance systems, the storage is chosen to provide fast writes, under the assumption that analysis will not occur online, meaning low-latency queries are not required. This is especially true for whole-system provenance like LPM, where large volumes of provenance data are collected. As an example, Table~\ref{tab:lpm-events} shows the volume of provenance data collected while compiling the kernel. What is needed is a system that can support the large volume of provenance data, while simultaneously supporting online analytics.

LPM-enabled systems can generate large volumes of data (between 2 and 4GB for an I/O intensive task, such as kernel compilation). For specifics on the volume of data generated, we refer the reader to~\cite{lpm} where the authors describe the storage overheads in great detail. Additionally, the types of data stored are described. In~\cite{lpm}, the authors explore the use of a local graph database running on the same system. While this local storage supports fast queries, it is limited to querying the provenance records for a single system. In contrast, storing provenance data in a centralized database supports more complex queries that can span multiple systems.

The configuration in this paper is as follows:
\begin{IEEEdescription}[\IEEEsetlabelwidth{Granularity}\IEEEusemathlabelsep]
  \item[Granularity] Whole-system
  \item[Collection] Linux Provenance Modules
  \item[Encoding] W3C PROV
  \item[Storage] Apache Accumulo
  \item[Analytics] Graph-traversal
  \item[Response] Passive
  \item[Security] Not addressed
\end{IEEEdescription}

Next, we detail the design of the ingest pipeline and the graph analysis developed to determine the inputs related to a specific output.

\section{Design} \label{sec:design}

\begin{figure}[t]
  \centering
  \includegraphics[width=3.15in]{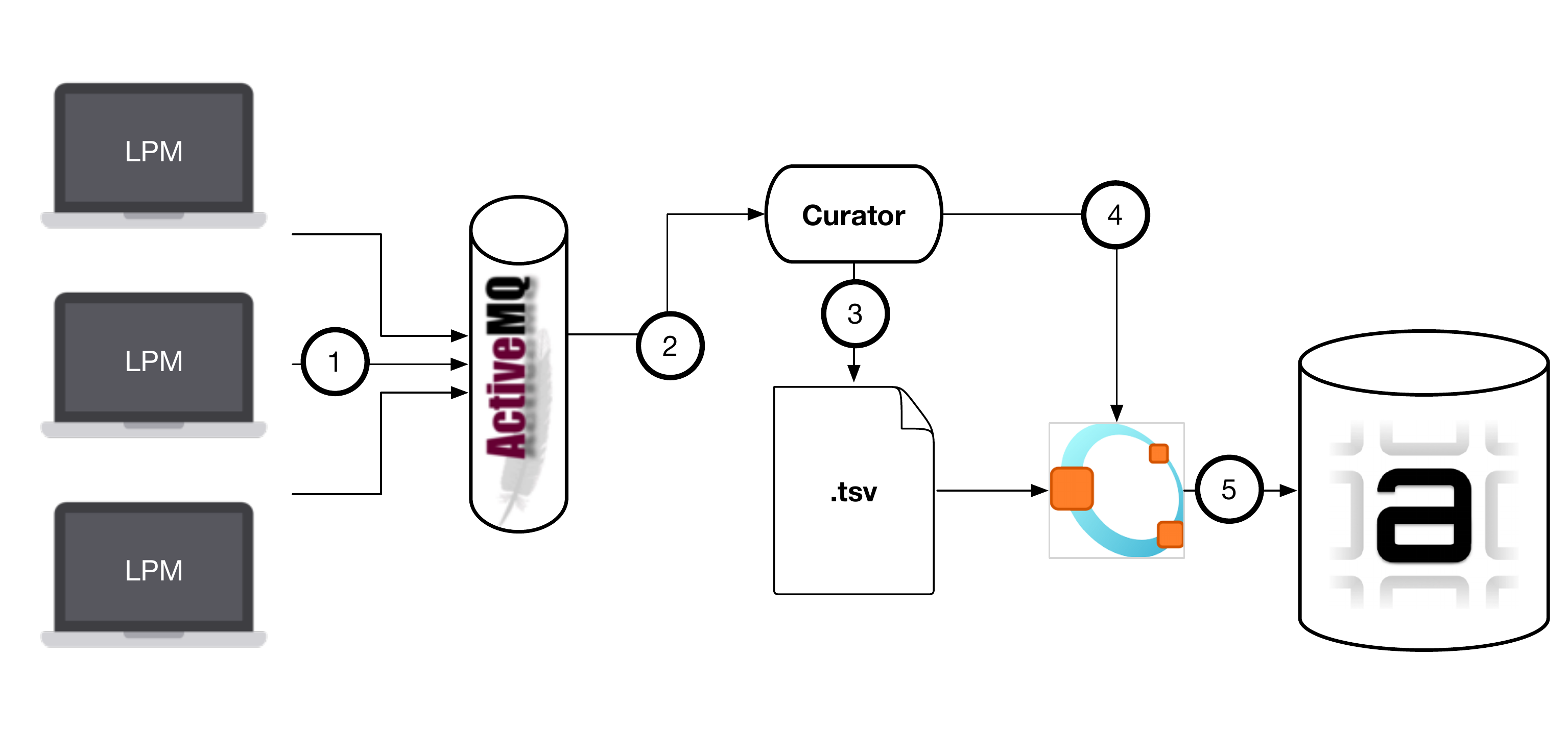}
  \caption{\label{fig:sys-arch} Our system uses LPM for provenance data collection, a custom application for encoding and storage of the provenance data from multiple systems (Curator). Curator relies on Matlab/GNU Octave and the D4M library storing the data in Accumulo.}
\end{figure}

Figure~\ref{fig:sys-arch} shows the architecture used to collect and store the provenance data. (1) Data is collected by the LPM-enabled kernel and sent to a central location via a message bus (ActiveMQ) by a daemon running on each LPM node. (2) The data is translated from LPM events into W3C PROV graphs by the Curator application\footnote{Internal application used as the daemon that collects provenance from a variety of sources.}, and then (3) encoded as tab-separated Accumulo entries based on the D4M schema. (4) After a configurable number of entries are process, Curator triggers Matlab/GNU Octave to process the TSV files. (5) Matlab/GNU Octave stores the entries in Accumulo using the open-source D4M library.

\begin{figure*}[t]
  \centering
  \includegraphics[width=5in]{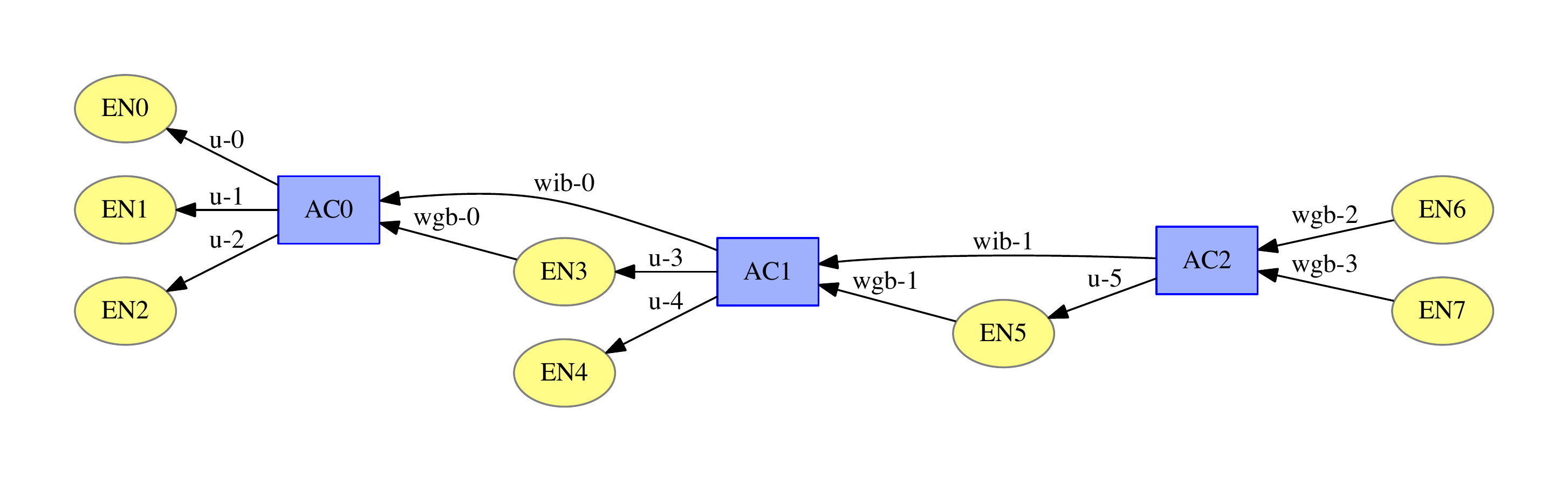}
  \caption{A simple provenance
    graph. Nodes that start with ``EN''
    are entities, or data. Nodes with ``AC'' indicate an activity (e.g. process)
    node. Directed edges in the graph follow the standard W3C PROV-DM model. In this simple example,
    the starting point of the graph is EN6 or EN7 which were generated
  by AC2 which was used by EN5, etc.}
\label{fig:simple_prov_graph} 
\end{figure*}

Each LPM event generates a single node and edge that must be encoded and stored in Accumulo. We rely on D4M~\cite{d4m} to encode the nodes and edges of the provenance graph. Data about nodes are stored in one table, while edges are stored in a separate table. Figure~\ref{fig:simple_prov_graph} shows a small provenance graph with entities (EN6 and EN7, for example) and activities (AC2, etc.). The definitions of the nodes and edges can be found in~\cite{prov-dm}.

\begin{lstlisting}[float,style=Matlab-editor,caption=Example Accumulo entries for nodes in Figure~\ref{fig:simple_prov_graph},frame=lines,label=lst:example-nodes,basicstyle=\footnotesize]
AC0 :type|PROV_ACTIVITY []    1
AC1 :type|PROV_ACTIVITY []    1
AC2 :type|PROV_ACTIVITY []    1
EN0 :type|PROV_ENTITY []    1
EN1 :type|PROV_ENTITY []    1
EN2 :type|PROV_ENTITY []    1
EN3 :type|PROV_ENTITY []    1
EN4 :type|PROV_ENTITY []    1
EN5 :type|PROV_ENTITY []    1
EN6 :type|PROV_ENTITY []    1
EN7 :type|PROV_ENTITY []    1
\end{lstlisting}

\begin{lstlisting}[float,style=Matlab-editor,caption=Example Accumulo entries for edges in Figure~\ref{fig:simple_prov_graph},frame=lines,label=lst:example-edges,basicstyle=\footnotesize]
...
wgb-2 :inNode|AC2 []    1
wgb-2 :inType|PROV_GENERATION|AC2 []    1
wgb-2 :outNode|EN6 []    1
wgb-2 :outType|PROV_GENERATION|EN6 []    1
wgb-2 :type|PROV_GENERATION []    1
wgb-3 :inNode|AC2 []    1
wgb-3 :inType|PROV_GENERATION|AC2 []    1
wgb-3 :outNode|EN7 []    1
wgb-3 :outType|PROV_GENERATION|EN7 []    1
wgb-3 :type|PROV_GENERATION []    1
...
\end{lstlisting}

Listings~\ref{lst:example-nodes} and~\ref{lst:example-edges} show examples of the data stored in Accumulo for the nodes and edges of the graph, respectively. The first column is the identifier of the node or edge. The second column stores nodes and edge attributes, such as the type (e.g.  \texttt{:type$|$PROV\_ACTIVITY}). For nodes shown in Listing~\ref{lst:example-nodes}, the node name is the first value, and the second value encodes the W3C PROV type of the node. Similarly for Listing~\ref{lst:example-edges}, each set of rows with the same name (e.g. wgb-2) encode the properties of the edge, such as source (the \texttt{:inNode} row), destination (the \texttt{:outNode}), and type (\texttt{:type}) of the edge. From the data in the tables, we are able to build provenance graphs.

\subsection{Provenance Graph Analysis}
 
Building the provenance graphs is an iterative process, that starts with identification of the starting node. For the LPM data this can be a socket, file, process, or user. In the example graph in Figure~\ref{fig:simple_prov_graph}, we will use ``EN6'' and ``EN7''.

\begin{enumerate}
\item Begin with starting node (``EN6'' and ``EN7'') and number of hops your would like to explore.
\item Find all edges connected to starting node(s). The ``\texttt{:type$|$PROV\_USAGE}'' are
  directed edges from an ``AC''' node (a process) to an ``EN'' (a
  file).  The ``\texttt{:type$|$PROV\_GENERATION}'' edges are directed edges from EN nodes
  (files) to AC nodes (processes). Other edges types, as defined in~\cite{prov-dm} are also encoded.
\item Find nodes connected via edges. These nodes are 1-hop distance
  from the starting node(s).
\item Continue process until depth (number of hops) is achieved. 
\end{enumerate}

For example, consider the code snippet and output shown in listings~\ref{lst:code}
and \ref{lst:output}, respectively. The code snippet shows a few
lines of MATLAB code used to traverse a provenance graph. The
first lines, set up a connection to the database system. The final
loop, finds nodes at the next
depth level that are connected to the nodes in the current level. For
figure~\ref{fig:simple_prov_graph}, the results of applying the
analytic to the first three depths will look like the contents of
listings~\ref{lst:output}  where \texttt{depthID} corresponds to the number of hops
from the starting node to the current node. 
Essentially, the algorithm performs a filtered breadth first search of the graph. The filter criteria can be edge or node attributes.

\begin{lstlisting}[float,style=Matlab-editor,caption=Sample of the code from the analytic,frame=lines,label=lst:code,basicstyle=\footnotesize]
%set binding to database
DB=DBsetupLLGrid('cyber-prov-db');

%set bindings to node and edge tables;
Tnode=DB('nodeTable');
Tedge=DB('edgeTable');

for i=1:depth
        Cedge=Row(Tedge(:,CatStr('outNode,','|', startFile)),:);
end

\end{lstlisting}

\begin{lstlisting}[float,style=Matlab-editor,caption={Example output},frame=lines,label=lst:output,basicstyle=\footnotesize]
(depthID|0,EN6,)     1,
(depthID|0,EN7,)     1,
(depthID|1,inNode|AC2,)     outNode|EN6,
(depthID|2,inNode|AC1,)     outNode|AC2,
(depthID|2,inNode|EN5,)     outNode|AC2,
(depthID|3,inNode|AC0,)     outNode|AC1,
(depthID|3,inNode|AC1,)     outNode|AC1,
(depthID|3,inNode|EN3,)     outNode|AC1,
(depthID|3,inNode|EN4,)     outNode|EN5,
...
\end{lstlisting}


\section{Evaluation} \label{sec:eval}

As highlighted in Table~\ref{tab:lpm-events}, the number of LPM events collected during a 38 minute kernel compile exceeds 4,000,000, giving a rate of 1,879 events per second. Each event generates one node and one edge that must be encoded and stored in the database, meaning any system that processes LPM data must support storing an average of 3,758 graph components per second.

For the experiments described in this section, we rely on a single Accumulo node running on a dual-socket Intel Xeon E5-2683 v3 system (28-cores total) with 256GB of RAM and 8TB of local storage. Each node is running Fedora 20, and the Accumulo node is running version 1.6. The nodes are connected via a 10GbE network. All times reported are from single runs.

In order to characterize the ingest rate of the pipeline shown in Figure~\ref{fig:sys-arch}, we used a random provenance graph generator that is part of the ProvToolbox library\footnote{\url{https://github.com/lucmoreau/ProvToolbox/tree/master/prov-generator}}. We generated graphs ranging from 2 to 524,288 nodes. Each graph was then ingested into the Accumulo database as described in Section~\ref{sec:design}. The graphs are designed to represent the sorts of graphs that are encountered when processing LPM events. The generator takes, as input, the number of nodes, and the maximum number of edges allowed per node. The tool then generates random nodes and edges that respect these inputs, and returns a valid provenance graph.\footnote{The reason we used random graphs is the ability to specify an arbitrary number of nodes and edges to better characterize the performance over a range of different inputs to the ingest pipeline.}

\begin{figure}[ht]
  \centering
  \includegraphics[width=2.85in]{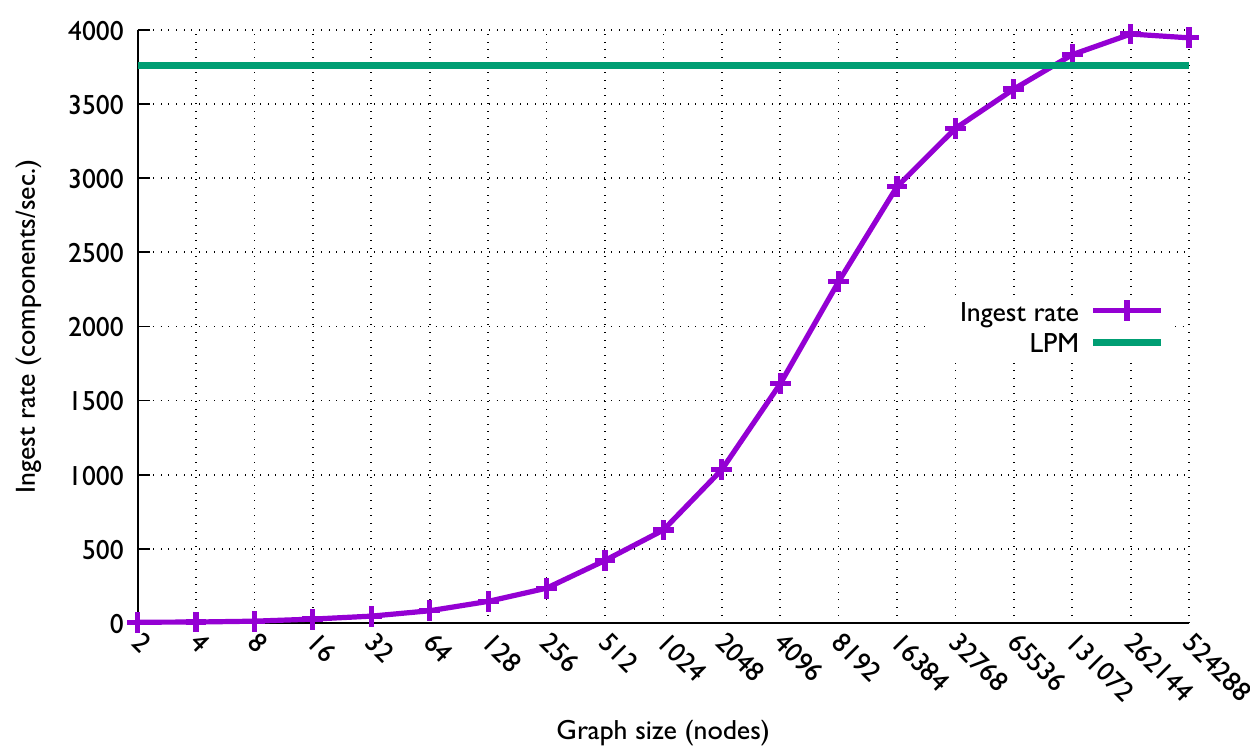}
  \caption{Graph size vs average component ingest rate}
  \label{fig:ingest}
\end{figure}

Figure~\ref{fig:ingest} shows the graph size versus average number of components (nodes and edges) ingested per second. From the graph, we see that batching the ingests is beneficial, with a peak of 3,970 components per second. This means that the ingest system can handle the data being generated by the LPM system. The ingest system relies on a single thread to write data to Accumulo, and one area of future work is to explore running multiple ingest threads to further increase the ingest rate to support multiple LPM nodes sending data to the database at once. It should also be noted that not all workloads generate the same volume of provenance data. Compute-intensive workloads will generate fewer events than I/O-intensive workloads\cite{lpm}.

\begin{figure}[ht]
  \centering
  \includegraphics[width=2.85in]{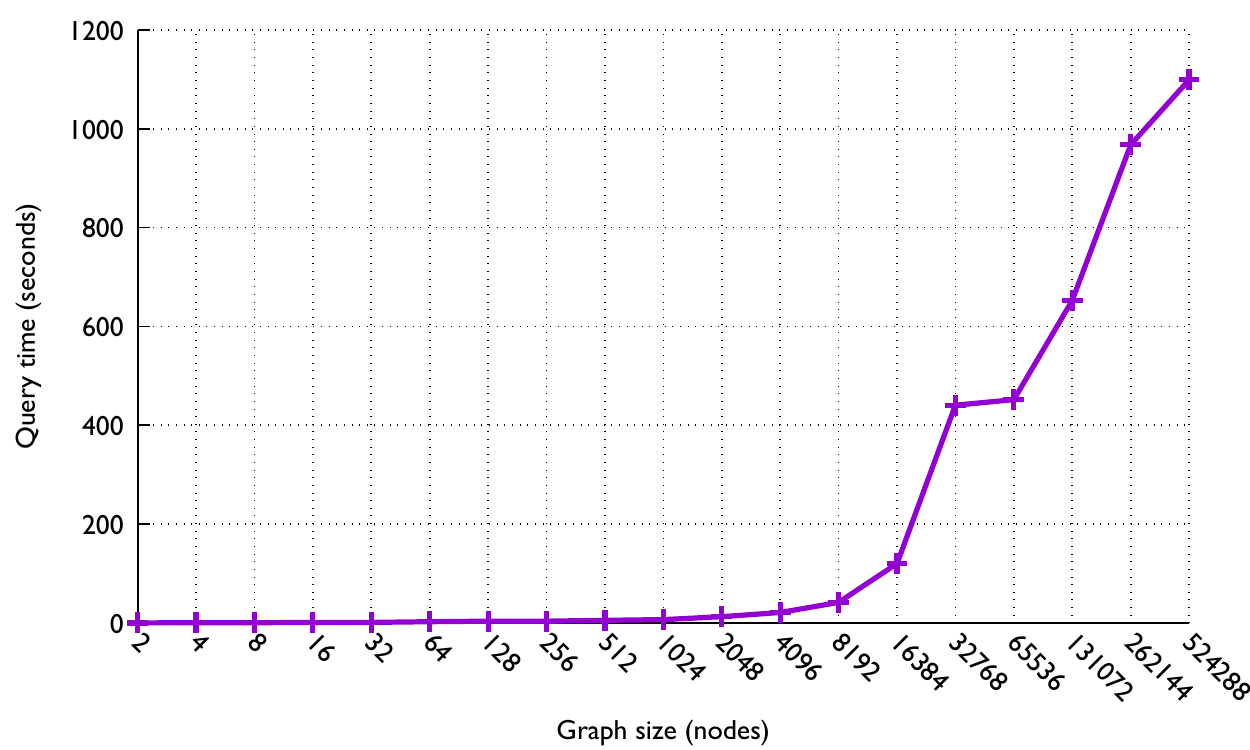}
  \caption{Graph size vs query time to build full graphs}
  \label{fig:query}
\end{figure}

Once the data is stored in Accumulo, we need to query the data to utilize it. In this example, the first step is to traverse the graph to identify the nodes and edges, starting from a given node of interest. From there, we perform a simple breadth-first search to build the graph, resulting in a listing similar to Listing~\ref{lst:output}. The query time for each graph size is shown in Figure~\ref{fig:query}. The graph query time is directly related to the depth of the graph, as the analytic must query the database for each hop in the graph. This brute force algorithm results in slow query times for large graphs. In Figure~\ref{fig:query}, the graph with ~32,000 nodes shows a larger than expected query time. This is a result of the random graph generator generating a graph that is deeper than subsequent graphs. In this case, the maximum depth was 20 levels, while the larger graph with ~64,000 nodes had a depth of only 13.

In the future, we are exploring ways to decrease this query time, and better characterize the ingest performance with a wider variety of real-world data. One possible way to reduce the query time is by leveraging Accumulo iterators to reduce the number of queries the client must make to retrieve the correct data. One server-side graph analysis library for Accumulo is Graphulo~\cite{graphulo}, which shows promise in helping to reduce the overhead of querying large graphs.


\section{Discussion} \label{sec:discussion}

\paragraph{Security} Security is a paramount concern for data provenance, especially if critical security decisions are going to be made. In this paper, we don't currently provide any protections for the data stored in Accumulo. There are several efforts underway to provide security for data stored in Accumulo. The first is the ``Computing on Masked Data'', or CMD, method~\cite{cmd,gadepally2015computing}, which allows data to be ``masked'' in Accumulo, and computations do not require data to be unmasked. The other approach is described in~\cite{pace-summit}, where the authors provide a client-side library that provides cryptographic protections for Accumulo data. In~\cite{secure-big-data-analytics}, the authors describe other methods for securing data that can be applied to this work.

\paragraph{Enhancements} While the ingest performance currently meets the needs for a single node producing LPM events under an I/O intensive workload, one area of future work will explore ways to increase the ingest performance such that a single node can support multiple clients at once. Furthermore, the analytics described in Section~\ref{sec:eval} are still too slow to support online use of the data in all but the smallest graphs. Exploring the use of Graphulo, and other server-side iterators, to reduce the query time is an area of future work.


\section{Conclusion} \label{sec:conclusion}

In this paper we present a system to support high-throughput storage of provenance data in Accumulo. We are just beginning to understand the needs of provenance in building resilient systems, and this work addresses one of the many challenges with integrating provenance into a system, namely the problem of storing large volumes of data. We have designed a system that supports high-throughput ingest of provenance data. Our system is able to sustain an ingest rate of approximately 4,000 components per second with a single Accumulo node. Additionally, we have started exploring the query interfaces for such a system, and identified one example query, namely traversing the graph to identify inputs related to a specific output. In the future, we are looking to reduce the query time to make our use of data provenance possible in online systems, as opposed to just offline analysis.


\section*{Acknowledgment}
The authors would like to thank the MIT Lincoln Laboratory
Supercomputing Center, Rob Cunningham
(Leader, Secure Resilient Systems and Technology, MIT Lincoln
Laboratory) and Dr. Marc Zissman (Associate Division Head, Cyber
Security and Information Sciences, MIT Lincoln Laboratory) for their support.

\begin{scriptsize}
\setlength{\parsep}{-1pt}\setlength{\itemsep}{0cm}\setlength{\topsep}{0cm}
\bibliographystyle{IEEEtran}
\bibliography{provenance-ingest}
\end{scriptsize}

\end{document}